\documentclass[11pt]{article}

\usepackage[utf8]{inputenc}
\usepackage[T1]{fontenc}
\usepackage{lmodern}
\usepackage[margin=1in]{geometry}
\usepackage{amsmath,amssymb}
\usepackage{graphicx}
\usepackage{booktabs}
\usepackage{tabularx}
\usepackage{array}
\usepackage{enumitem}
\usepackage{xcolor}
\usepackage{xurl}
\usepackage[numbers,square,sort&compress]{natbib}
\usepackage{tikz}
\usetikzlibrary{arrows.meta,positioning,shapes.geometric,fit,calc,backgrounds}
\usepackage{authblk}
\usepackage[colorlinks=true,linkcolor=blue,citecolor=blue,urlcolor=blue]{hyperref}
\usepackage{doi}

\newcolumntype{Y}{>{\raggedright\arraybackslash}X}
\newcommand{\keywords}[1]{\vspace{0.5em}\noindent\textbf{Keywords:} #1}

\title{Building a Regional Data-Centric Materials Science Ecosystem for Processing-Rich Materials Innovation in the Great Plains}
\author[1,*]{D.-M. Mei}
\author[18]{K. Acharya}
\author[2]{C. M. Adhikari}
\author[1]{M. Adhikari}
\author[11]{S. Aryal}
\author[3]{B. V. Benson}
\author[11]{K. Bhatta}
\author[1]{S. Bhattarai}
\author[1]{N. Budhathoki}
\author[28]{A. M. Castillo}
\author[4]{D. Chakraborty}
\author[1]{S. Chhetri}
\author[5]{S. Choudhury}
\author[6]{T. A. Chowdhury}
\author[8]{R. D. Cruz}
\author[7]{B. Cui}
\author[11]{S. Dhital}
\author[1]{K.-M. Dong}
\author[9]{R. Gapuz}
\author[10]{A. Ghasemi}
\author[11]{E. Z. Gnimpieba}
\author[11]{B. D. S. Gurung}
\author[12]{H. A. Hashim}
\author[13]{R. I. Harry}
\author[14]{K.-E. Hasin}
\author[29]{M. K. Hassanzadeh}
\author[1]{M. K. Jha}
\author[1]{D. Kim}
\author[6]{K.-C. Kong}
\author[4]{B. Lama}
\author[15]{A. Mahat}
\author[11]{N. Maharjan}
\author[26]{A. Majeed}
\author[1]{J. Mammo}
\author[16]{M. M. Masud}
\author[17]{K. S. Moore}
\author[30]{T. Mukherjee}
\author[10]{A. Nawaz}
\author[18]{H. Oli}
\author[1]{S. A. Panamaldeniya}
\author[1]{L. Pandey}
\author[27]{R. Pandey}
\author[19]{Z. Peng}
\author[1]{A. Prem}
\author[11]{M. M. Rana}
\author[20]{K. Rana Magar}
\author[21]{R. Rizk}
\author[22]{C. S. Tadi}
\author[11]{L.-W. Wang}
\author[1]{Y. Yang}
\author[23]{G.-L. Yin}
\author[23]{C.-X. Yu}
\author[24]{D. Zeng}
\author[25]{M. Zhou}
\author[6]{Q. Zhou}

\affil[1]{Department of Physics, University of South Dakota, Vermillion, SD 57069, USA}
\affil[2]{Department of Chemistry, Physics and Materials Science, Fayetteville State University, Fayetteville, NC 28301, USA}
\affil[3]{PROMISE Lab, Sanford Research, Sioux Falls, SD 57104, USA}
\affil[4]{Department of Physics and Astronomy, University of Nebraska at Kearney, Kearney, NE 68849, USA}
\affil[5]{Department of Mechanical Engineering, University of Mississippi, University, MS 38677, USA}
\affil[6]{Department of Physics and Astronomy, University of Kansas, Lawrence, KS 66045, USA}
\affil[7]{Department of Mechanical and Materials Engineering, University of Nebraska--Lincoln, Lincoln, NE 68588, USA}
\affil[8]{Tiospa Zina Tribal School, Agency Village, SD 57262, USA}
\affil[9]{Dakota State University, Madison, SD 57042, USA}
\affil[10]{Department of Physics, University of Cincinnati, Cincinnati, OH 45221, USA}
\affil[11]{University of South Dakota, Vermillion, SD 57069, USA}
\affil[12]{Department of Physics, University of Central Florida, Orlando, FL 32816, USA}
\affil[13]{Department of Engineering and Applied Sciences, RAND Corporation, Santa Monica, CA 90401, USA}
\affil[14]{University of California, Merced, CA 95343, USA}
\affil[15]{School of Computing, Wichita State University, Wichita, KS 67260, USA}
\affil[16]{University of Illinois Chicago, Chicago, IL 60607, USA}
\affil[17]{STEP Lab, Massachusetts Institute of Technology, Cambridge, MA 02139, USA}
\affil[18]{Department of Physics, South Dakota School of Mines and Technology, Rapid City, SD 57701, USA}
\affil[19]{Department of Computer Science, University of Montana, Missoula, MT 59812, USA}
\affil[20]{Georgia State University, Atlanta, GA 30303, USA}
\affil[21]{Department of Computer Science, University of South Dakota, Vermillion, SD 57069, USA}
\affil[22]{College of Science and Engineering, Saint Cloud State University, St. Cloud, MN 56301, USA}
\affil[23]{Department of Agricultural and Biosystems Engineering, Iowa State University, Ames, IA 50011, USA}
\affil[24]{Information Systems, Dakota State University, Madison, SD 57042, USA}
\affil[25]{Department of Chemical and Materials Engineering, New Mexico State University, Las Cruces, NM 88003, USA}
\affil[26]{Department of Mechanical Engineering, Villanova University, Villanova, PA 19085, USA}
\affil[27]{Department of Computer Science, University of Kentucky, Lexington, KY 40506, USA}
\affil[28]{Lower Brule Schools, Lower Brule, SD 57548, USA}
\affil[29]{Department of Physics, Virginia Tech University, Blacksburg, VA 24061, USA}
\affil[30]{Department of Mechanical Engineering, Iowa State University, Ames, IA 50011, USA}
\affil[*]{Corresponding author: \href{mailto:Dongming.Mei@usd.edu}{Dongming.Mei@usd.edu}}
\date{\today}

\begin{document}

\maketitle
\vspace{-1.5em}

\begin{abstract}
Data-centric materials science is changing how materials are discovered, optimized, manufactured, and qualified, yet many deployment-limiting materials problems still depend on experimental, processing-rich, device-level, and field-relevant data that are difficult to capture in conventional materials databases. This perspective argues that the Great Plains and adjacent interior research corridor can make a distinctive national contribution by organizing distributed experimental assets into a trusted regional materials-data ecosystem. The proposed model emphasizes FAIR metadata, provenance, persistent sample identifiers, uncertainty-aware modeling, semi-closed-loop workflows, stackable workforce training, and tiered governance for academic, public, controlled-access, and industry-protected data. We identify five coupled barriers---fragmented data, weak algorithm--laboratory translation, uneven access to cyberinfrastructure and technical staff, workforce gaps at the materials--data interface, and insufficient incentives for sharing and reuse---and propose a staged roadmap for addressing them. A high-purity germanium pilot illustrates how regional strengths can be converted into reusable datasets, benchmark models, trained personnel, and decision-improving workflows. The broader message is that regional leadership in data-centric materials science will depend less on geographic concentration than on trustworthy data practices, interoperable infrastructure, cross-trained people, and application-driven materials challenges.
\end{abstract}


\section{Introduction}

Materials discovery has historically advanced through scientific intuition, mechanistic modeling, incremental experimentation, and engineering optimization. This paradigm has produced transformative technologies in electronics, energy, structural materials, catalysis, biomedical devices, and manufacturing. It is increasingly strained, however, when design spaces are high-dimensional, processing--structure--property relationships are nonlinear, or qualification requires repeated cycles of synthesis, characterization, reliability testing, and scale-up. These challenges are especially acute for emerging materials systems whose performance depends not only on composition and crystal structure, but also on defects, interfaces, microstructure, processing history, device architecture, and operating environment.

Data-centric materials science addresses these bottlenecks by treating data, algorithms, experimental infrastructure, and domain expertise as coequal elements of materials innovation~\cite{Curtarolo2013,Agrawal2016,Butler2018}. In this framework, data are not merely the final output of experiments or simulations. They are reusable scientific assets that support prediction, uncertainty quantification, experiment selection, model validation, and cross-laboratory learning. The field has been accelerated by open materials databases, high-throughput density-functional-theory workflows, automated synthesis and characterization platforms, machine-learning toolkits, and FAIR data principles~\cite{Jain2013,Ward2018,Choudhary2020,Draxl2019,Wilkinson2016}. National efforts such as the Materials Genome Initiative (MGI) and NSF Materials Innovation Platforms further emphasize the need to unify materials innovation infrastructure, harness materials data, and train a connected materials research-and-development workforce~\cite{MGI2021,NSFMIP2025}.

A critical gap remains. Many national materials-data resources are strongest for computed structures, idealized properties, and highly curated benchmark datasets, whereas many deployment-limiting problems depend on smaller but information-rich experimental datasets. These include processing histories, instrument metadata, calibration records, uncertainty estimates, failed experiments, environmental exposure, device architecture, lifetime behavior, and manufacturing constraints. This gap creates an opportunity for regions with strong experimental, manufacturing, agricultural, energy, and device-test capabilities.

The distinctive contribution of this perspective is not to propose another stand-alone national database or to duplicate existing high-throughput computational resources. Instead, it develops a regional operating model for converting distributed experimental laboratories into a coordinated source of processing-rich, device-relevant, uncertainty-aware materials data. In this model, the highest-value products are not only publications, but also reusable data packets, metadata schemas, sample-ID protocols, model cards, benchmark tasks, training modules, and decision records that can improve the next experiment.

The motivation for this perspective grew out of the Workshop for AI-Powered Materials Discovery at Great Plains~\cite{AIMaterialsWorkshop2025}, held at the University of South Dakota from June 22--25, 2025, with support from the National Science Foundation EPSCoR Workshop Program. The workshop was well attended, bringing together more than 200 participants and highlighting both the scientific opportunity and the coordination challenge facing the region. Discussions at the workshop made clear that the Great Plains contains substantial but distributed strengths in materials synthesis, characterization, computation, cyberinfrastructure, workforce training, and use-inspired research. They also revealed a shared need for common data practices, interoperable workflows, cross-institutional training, and stronger pathways for translating AI-enabled materials discovery into deployable technologies. This paper responds to that regional momentum by framing a practical roadmap for building a trusted, data-centric materials science ecosystem across the Great Plains and adjacent interior research corridor.

For this paper, the term \emph{Great Plains and adjacent interior research corridor} is used functionally rather than as a strict geographic label. The core Great Plains states include North Dakota, South Dakota, Nebraska, Kansas, and Oklahoma. Adjacent interior partners include Iowa, Montana, Wyoming, Colorado, New Mexico, and Texas, whose universities, national-laboratory connections, cyberinfrastructure, energy systems, agricultural technologies, manufacturing sectors, and user-facility access points can strengthen a regional materials-data network. This definition is intended to identify a practical collaboration corridor organized around shared materials challenges, workforce needs, infrastructure complementarities, and cross-state research opportunities.

The inclusion criteria are therefore functional rather than political: a partner strengthens the corridor if it contributes complementary experimental capabilities, user-facility access, cyberinfrastructure, workforce pathways, industry pull, or deployable use cases that cannot be supplied by one institution alone. This framing also keeps the argument bounded. The corridor is broad enough to support modern materials-data workflows, but narrow enough to organize around common EPSCoR-relevant workforce needs, rural and regional innovation priorities, and shared application domains.

\begin{figure}[htp!]
    \centering
    \includegraphics[width=0.92\linewidth]{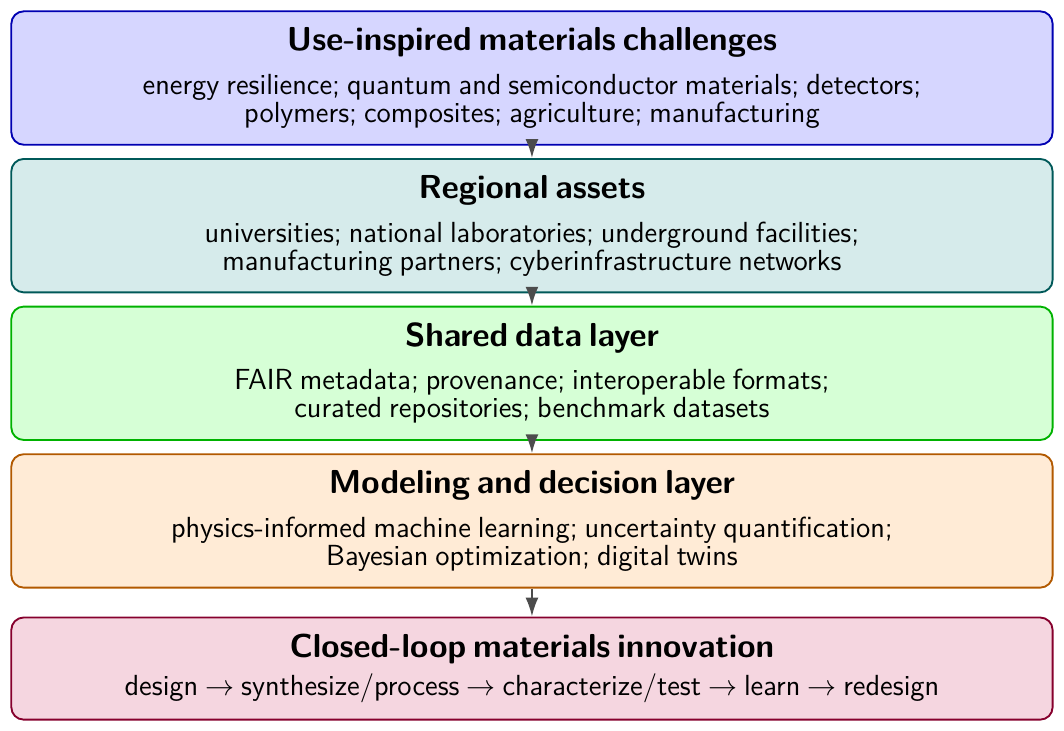}
    \caption{Conceptual architecture for a regional data-centric materials science ecosystem. The regional strategy connects use-inspired challenges, distributed assets, shared data, modeling, and closed-loop experimental learning.}
    \label{fig:regional_data_centric_architecture}
\end{figure}

The central thesis of this perspective is that the Great Plains can become a visible national contributor to data-centric materials science by specializing in experimental, processing-rich, application-driven datasets and by connecting those datasets to shared cyberinfrastructure, workforce training, and semi-closed-loop laboratory workflows. The region need not reproduce the scale or density of established coastal ecosystems. Instead, it can build leadership by integrating distinctive strengths in high-purity semiconductors and detector materials, quantum and cryogenic devices, energy and environmental resilience, polymers and composites, agriculture, manufacturing, underground test environments, health technologies, and cyberinfrastructure-enabled collaboration. Figure~\ref{fig:regional_data_centric_architecture} summarizes this strategy.

This paper makes three contributions. First, it provides a regional landscape analysis that identifies the assets most relevant to data-centric materials science in the Great Plains and adjacent interior corridor. Second, it analyzes strategic barriers that currently prevent distributed capabilities from becoming a coherent materials-data ecosystem. Third, it proposes a staged roadmap and a concrete pilot workflow that convert broad regional aspirations into actionable deliverables: metadata templates, sample identifiers, reusable datasets, benchmark models, training modules, governance policies, and measurable translation metrics. The intended audience includes materials researchers, cyberinfrastructure developers, university and state research leaders, federal program officers, and industry partners interested in building distributed, data-enabled materials innovation ecosystems.

\section{From Materials Informatics to Closed-Loop Materials Innovation}

\begin{figure}[htp!]
    \centering
    \includegraphics[width=0.88\linewidth]{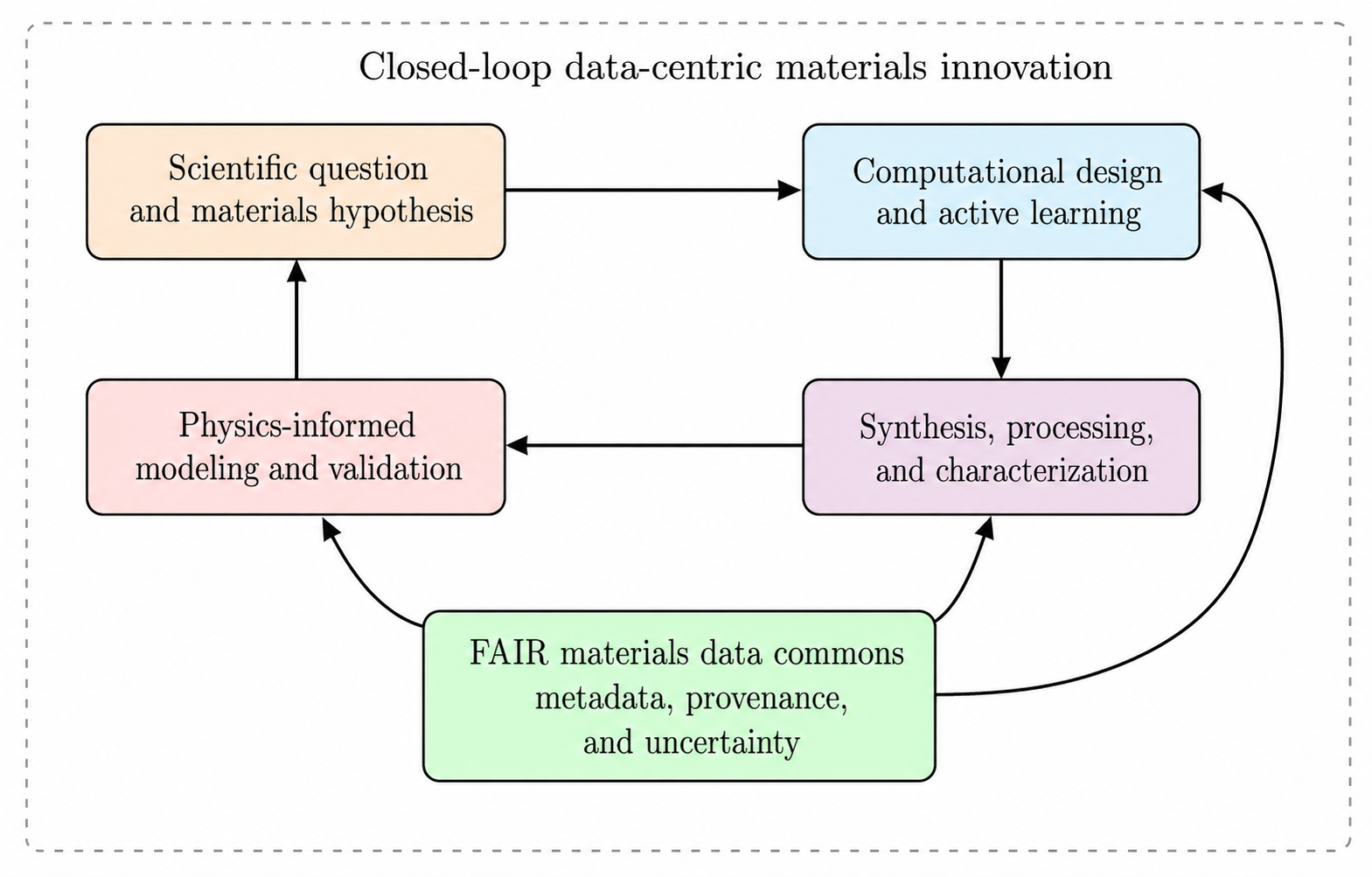}
    \caption{Conceptual closed-loop workflow for data-centric materials innovation. The loop links scientific questions and materials hypotheses, computational design and active learning, synthesis and characterization, and physics-informed modeling and validation. A FAIR materials data commons serves as an active infrastructure layer supporting metadata, provenance, and uncertainty-aware data exchange across the workflow.}
    \label{fig:closed_loop_data_centric}
\end{figure}

Data-centric materials science is often used interchangeably with materials informatics, but the two terms should be distinguished. Materials informatics usually refers to the extraction of knowledge from materials data using statistical learning, data mining, and machine learning. Data-centric materials science is broader: it includes the governance, instrumentation, metadata, repositories, software, human training, and organizational structures needed to make materials data useful across the full discovery-to-deployment pipeline. Figure~\ref{fig:closed_loop_data_centric} summarizes this conceptual loop.

A mature closed-loop workflow captures raw, processed, and derived data with provenance; combines physics-based and data-driven models; uses uncertainty to prioritize measurements; records unsuccessful experiments and boundary conditions; and embeds human expertise through constraints, priors, mechanistic features, and interpretation. These requirements are practical, not merely philosophical: without them, models may fit existing data but fail to guide the next useful experiment.

Autonomous and semi-autonomous laboratories show how models, robotics, active learning, and rapid characterization can reduce the number of experiments needed to identify promising candidates~\cite{Tabor2018,Szymanski2023,Hung2024}. Most regional laboratories, however, will first need \emph{automation-ready} and \emph{data-ready} workflows: standardized recipes, electronic laboratory notebooks, instrument APIs, shared metadata schemas, and model-guided experimental planning. These lower-barrier steps create the foundation for higher levels of autonomy while remaining realistic for heterogeneous equipment and limited technical staff.


\section{Current Regional Landscape}
\label{sec:regional_landscape}

The Great Plains and adjacent interior research corridor contains a distributed ecosystem that can support data-centric materials science. Its assets are spread across land-grant universities, emerging research universities, specialized laboratories, national-laboratory partnerships, agricultural and energy industries, advanced manufacturing, underground research infrastructure, and regional cyberinfrastructure networks. This geography creates coordination challenges, but it is also well matched to modern data-intensive research, where value can be created by connecting distributed instruments, people, and use-inspired testbeds through common metadata, persistent identifiers, reproducible workflows, and shared training~\cite{Wilkinson2016,MGI2021,NSFCSSI2022}.

\begin{figure}[htp!]
    \centering
    \includegraphics[width=0.92\linewidth]{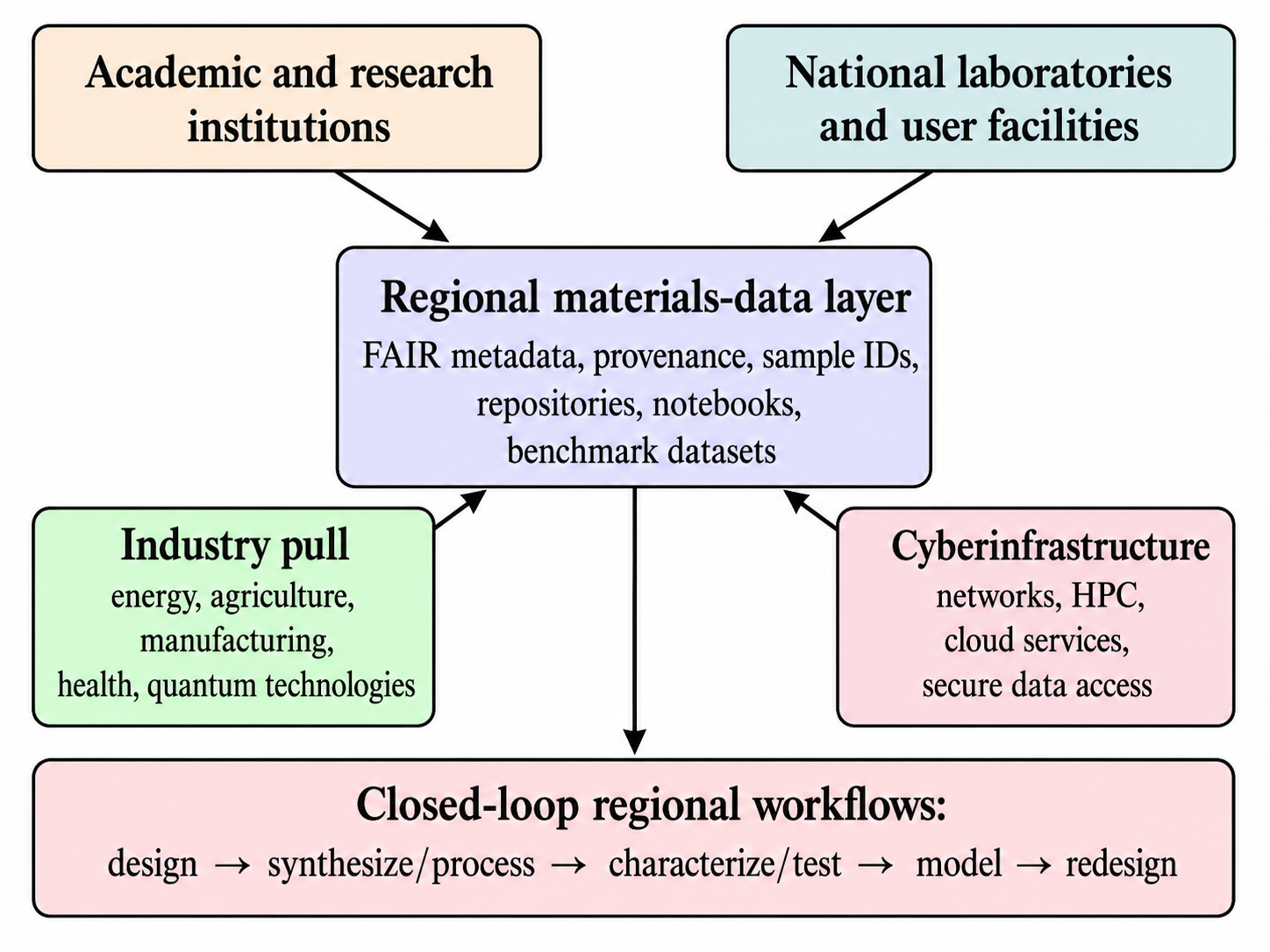}
    \caption{Networked landscape for data-centric materials science in the Great Plains and adjacent interior research corridor. A shared materials-data layer connects academic, national-laboratory, industry, and cyberinfrastructure assets to closed-loop regional workflows.}
    \label{fig:regional_landscape_network}
\end{figure}

A data-centric view changes how the regional landscape should be evaluated. The relevant question is not only whether individual institutions have strong materials synthesis, characterization, computation, or device-development capabilities. The strategic question is whether those capabilities can be connected into a network that produces reusable data and better materials decisions. Figure~\ref{fig:regional_landscape_network} shows this networked view.

\subsection{Regional Definition and Scope}
\label{subsec:regional_definition_scope}

The regional boundary used here is intentionally functional. The core Great Plains states provide the agricultural, energy, rural-manufacturing, workforce, and emerging-research-institution context. Adjacent interior states add complementary assets, including large materials programs, national-laboratory connections, advanced characterization capabilities, cyberinfrastructure, and industrial partners. The paper therefore uses three nested categories. The first is the \emph{core Great Plains}, where the need for regional coordination and workforce retention is most acute. The second is the \emph{adjacent interior corridor}, which supplies complementary research capacity and user-facility access. The third is the \emph{operational national network}, including DOE user facilities and national data resources that regional teams can access but do not need to reproduce locally.

This scope avoids two weaknesses. A definition that is too narrow would understate the collaborations required for modern materials science. A definition that is too broad would dilute the regional argument. The functional corridor definition is therefore used to identify assets that can contribute to common materials-data infrastructure, shared workforce programs, and application-driven pilot workflows.

This distinction also applies to authorship. The author list reflects participation in a broader workshop and collaboration network associated with this agenda, rather than a strict census of institutions inside the functional corridor. Accordingly, some coauthors are based outside the corridor, and some corridor states are represented through regional assets, partners, or intended participation rather than current coauthorship.

\subsection{Asset Categories and Strategic Opportunities}
\label{subsec:asset_categories}

Table~\ref{tab:regional_assets_opportunities} summarizes the major asset categories, representative examples, data-centric relevance, and strategic gaps. The table is not intended to be an exhaustive inventory of every institution in the region. Rather, it provides an analytical framework for identifying where a regional materials-data ecosystem can add value.

\begin{table}[htp!]
\centering
\small
\caption{Representative regional assets and data-centric opportunities. The purpose of the table is to move from a descriptive list of institutions to an analytical view of how regional capabilities can generate reusable materials data and decision-improving workflows.}
\label{tab:regional_assets_opportunities}
\begin{tabularx}{\linewidth}{p{0.18\linewidth}Y Y Y}
\toprule
\textbf{Asset category} & \textbf{Representative examples} & \textbf{Data-centric relevance} & \textbf{Gap or opportunity} \\
\midrule
High-purity semiconductors, detectors, and quantum materials & USD high-purity Ge crystal growth and detector development; cryogenic and device-test partners; university semiconductor and quantum-materials groups & Processing-rich data linking purification, crystal growth, contacts, cryogenic operation, leakage, capacitance, energy resolution, and qubit-relevant device metrics & Develop a growth-to-device metadata model, persistent sample IDs, and benchmark datasets for yield, stability, and device qualification \\
\addlinespace
Underground and low-background environments & Sanford Underground Research Facility (SURF); low-background counting, radiopurity, cosmogenic-activation, and rare-event detector programs & Environment-specific data linking material origin, underground storage or processing, exposure history, radioassay, background rates, shielding, and long-duration stability & Add low-background and exposure-history metadata to regional templates so rare-event and quantum-device datasets capture underground-environment advantages \\
\addlinespace
Critical, energy, and environmental materials & Ames National Laboratory; university energy-materials programs; regional energy and infrastructure sectors & Data linking synthesis, processing, degradation, field exposure, corrosion, cycling, and lifetime performance & Build field-relevant datasets and uncertainty-aware models for deployment decisions, not only discovery screening \\
\addlinespace
Agriculture, manufacturing, polymers, and composites & Land-grant universities; manufacturing partners; additive-manufacturing and composite laboratories & Data linking formulation, feedstock, processing, microstructure, mechanical performance, wear, humidity, ultraviolet exposure, and failure modes & Create industry-compatible data templates and protected access tiers for proprietary processing and performance data \\
\addlinespace
Biomaterials and health technologies & University biomedical, sensor, polymer, imaging, and radiation-detection groups & Multimodal data connecting surface chemistry, biological assays, sterilization, sensor response, imaging, and regulatory constraints & Standardize assay metadata, model documentation, and governance for sensitive or regulated data \\
\addlinespace
Cyberinfrastructure and workforce networks & Great Plains Network; institutional HPC/cloud resources; shared notebooks; regional training programs & Enables distributed storage, controlled access, reproducible analysis, cross-institution training, and workflow reuse & Provide shared data stewards, common templates, containerized workflows, and stackable training modules \\
\bottomrule
\end{tabularx}
\end{table}

\subsection{Academic and Research Institutions}
\label{subsec:academic_research_institutions}

The region includes universities with strengths in materials synthesis, characterization, computation, semiconductor and quantum materials, polymers and composites, nanotechnology, energy materials, photonics, biomedical materials, and advanced manufacturing. Representative examples include the University of Nebraska--Lincoln and its Nebraska Center for Materials and Nanoscience~\cite{UNLNCMN}; South Dakota Mines, with research strengths in advanced materials and manufacturing, underground science and engineering, critical and sustainable resources, and resilient cyberinfrastructure~\cite{SDMinesResearch}; the University of South Dakota, with distinctive capabilities in high-purity germanium crystal growth, detector development, and emerging AI-guided optimization of detector-grade crystal production~\cite{Wang2015,USDGeResearch2025}; the University of Colorado Boulder, whose materials science and engineering activities span sustainability, renewable energy, polymers, nanotechnology, electronics, photonics, and advanced computing~\cite{CUBoulderMSE}; the University of Texas at Austin and the Texas Materials Institute~\cite{UTAustinTMI}; and the University of New Mexico Center for High Technology Materials~\cite{UNMCHTM}.

These institutions are complemented by land-grant and regional universities with strong connections to agriculture, energy, environmental science, manufacturing, and workforce development. This combination matters because data-centric materials science requires more than high-throughput computation. It requires realistic materials-processing histories, operating environments, degradation data, and end-user constraints. Regional universities and industry-facing laboratories can therefore contribute experimental datasets that are difficult to generate in purely computational or highly centralized settings.

A distinctive feature of the region is the presence of specialized experimental environments and materials platforms. The Sanford Underground Research Facility (SURF) in South Dakota provides an internationally visible underground environment for low-background physics, rare-event searches, and detector development~\cite{Heise2015,SURF}. For materials science, such an environment is relevant to radiopurity, cosmogenic activation, low-background fabrication, cryogenic operation, and long-duration stability studies. These specialized environments strengthen the case for a regional strategy focused on experimental, processing-rich, and qualification-relevant data.

\subsection{National Laboratory and User-Facility Assets}
\label{subsec:national_labs_user_facilities}

Ames National Laboratory in Iowa is a major Department of Energy laboratory with longstanding strengths in materials science, rare-earth elements, critical materials, synthesis, and materials processing. Its Critical Materials Innovation Hub focuses on accelerating scientific and technological solutions for resilient supply chains of rare-earth metals and other materials important to energy technologies~\cite{AmesLabCMI,DOE_CMI2024}. For a regional materials-data strategy, Ames can provide specialized synthesis and characterization capabilities, help define metadata practices consistent with national expectations, connect regional teams to critical-materials priorities, and serve as a bridge to federal programs.

The region should also treat national user facilities outside its formal geography as operational assets. The DOE Office of Science maintains a national network of user facilities that provide light sources, neutron sources, electron-beam characterization, nanoscale science centers, accelerators, environmental facilities, and high-performance computing resources~\cite{DOEUserFacilities,DOEBESUserFacilities}. The strategic question is not whether every instrument exists within the Great Plains. It is whether regional teams can access distributed instruments, capture the resulting data in interoperable formats, and integrate those data into reusable workflows. Before experiments are performed, sample identifiers, processing histories, measurement metadata, calibration records, uncertainty estimates, and data-use conditions should be defined.

\subsection{Industry Pull and Cyberinfrastructure}
\label{subsec:industry_ci}

The Great Plains economy creates strong application pull. Energy industries need materials for batteries, catalysts, grid storage, geothermal systems, hydrogen technologies, carbon management, corrosion-resistant infrastructure, and low-emission industrial processes. Agricultural and food systems need durable coatings, polymer packaging, cold-chain materials, sensors, membranes, and precision-agriculture materials. Aerospace, defense-related, and advanced-manufacturing sectors need lightweight composites, high-temperature alloys, additive-manufacturing feedstocks, qualification protocols, and nondestructive evaluation methods. Health-technology applications require biomaterials, antimicrobial surfaces, radiation detectors, imaging materials, and sensor platforms. Emerging quantum and semiconductor technologies require high-purity crystals, low-defect interfaces, cryogenic-compatible contacts, reproducible nanofabrication, and reliable device-level characterization.

These use cases share a common bottleneck: deployment depends on reliable data across composition, processing, structure, property, performance, and lifetime. A coating dataset that records surface preparation, humidity, curing conditions, substrate condition, exposure history, failure mode, and uncertainty may be more valuable than a larger dataset that records only nominal composition and one property value. Similarly, a semiconductor or detector-materials dataset that includes impurity levels, thermal history, contact processing, leakage current, capacitance, energy resolution, and cryogenic operating conditions can support more useful prediction than a dataset limited to composition or crystal structure. This is where the region can make a distinctive contribution to national materials-data infrastructure.

Because the region is geographically distributed, cyberinfrastructure is a central enabling technology. The Great Plains Network provides a model for regional connectivity, shared research cyberinfrastructure, and inter-institutional collaboration~\cite{GreatPlainsNetwork}. NSF's Cyberinfrastructure for Sustained Scientific Innovation program similarly emphasizes integrated cyberinfrastructure services, measurable outcomes, and sustainable research communities~\cite{NSFCSSI2022}. A regional strategy should therefore include shared metadata templates, persistent identifiers, controlled-access storage, cloud-accessible notebooks, containerized workflows, and training materials that allow smaller institutions to participate without duplicating full-scale infrastructure locally.

\subsection{Landscape Assessment}
\label{subsec:landscape_assessment}

The region has domain depth, application breadth, and distinctive experimental environments, but it lacks a coordinated materials-data framework. Without that framework, datasets remain fragmented, metadata remain inconsistent, and machine-learning efforts remain difficult to reproduce, compare, or scale. The most important opportunity is therefore not simply to add more isolated projects. It is to build an active regional data layer that supports model training, uncertainty quantification, experiment selection, benchmark development, technology translation, and workforce education. Table~\ref{tab:regional_assets_opportunities} and Figure~\ref{fig:regional_landscape_network} together define the strategic premise for the remainder of the paper: regional leadership should be built around reusable experimental data and decision-improving workflows.

\section{Strategic Gaps and Barriers}
\label{sec:strategic_gaps}

\begin{figure}[htp!]
    \centering
    \includegraphics[width=0.98\textwidth]{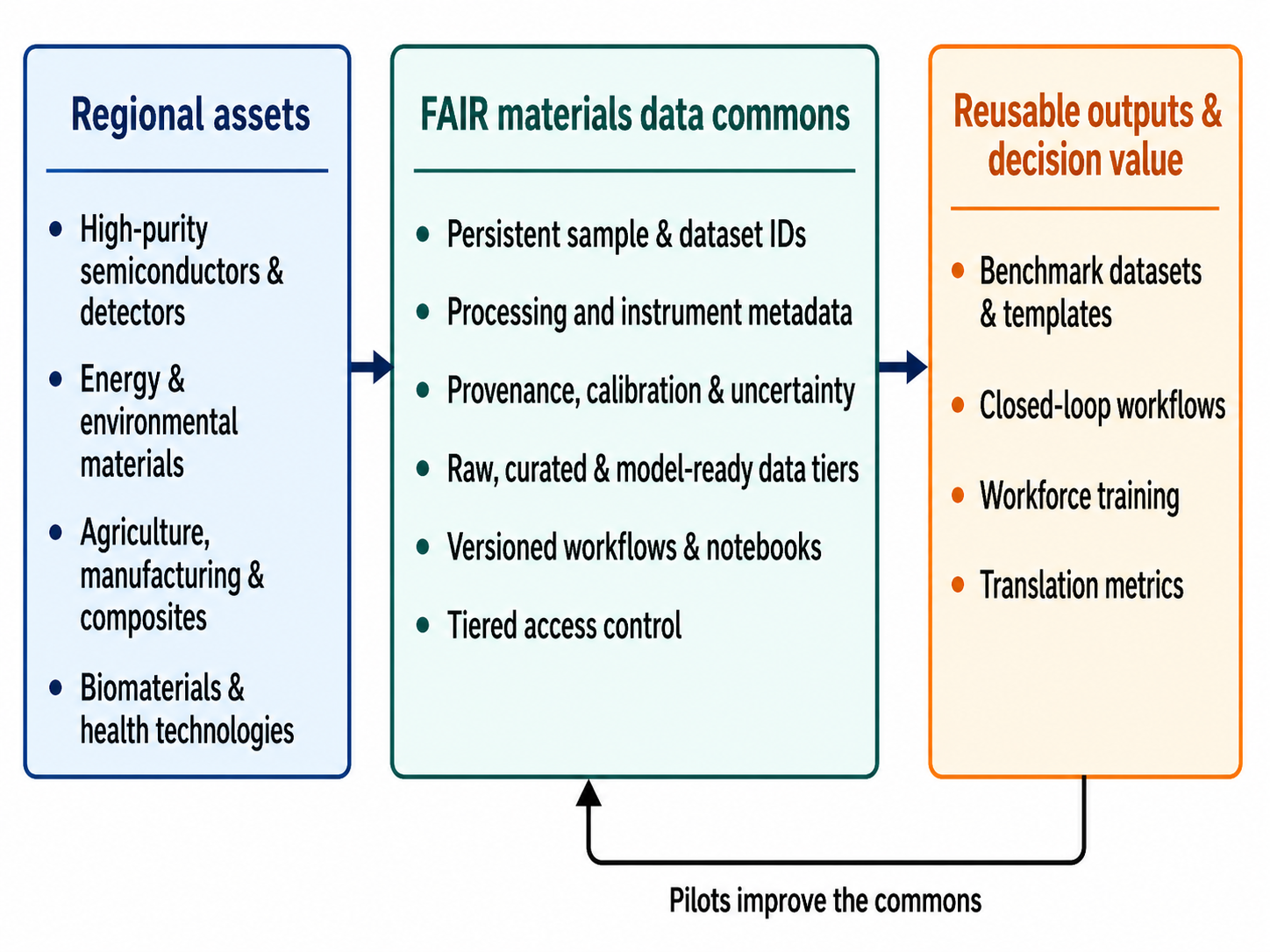}
    \caption{Regional asset-to-data-commons strategy. Distributed assets become reusable, decision-ready outputs when routed through a FAIR-aligned commons with identifiers, provenance, workflows, access tiers, and benchmark resources.}
    \label{fig:regional_asset_commons}
\end{figure}

The Great Plains and adjacent interior corridor has substantial materials assets, but assets alone will not produce regional leadership. The limiting factors are coupled: fragmented data reduce model quality; weak algorithm--laboratory translation reduces trust; uneven access to cyberinfrastructure and technical staff limits participation; workforce gaps slow adoption; and weak incentives keep valuable data in local notebooks, instrument folders, or proprietary archives. Figure~\ref{fig:regional_asset_commons} illustrates how regional assets can be converted into reusable, decision-ready data products through a shared FAIR-aligned materials data commons, while Table~\ref{tab:barriers_interventions} summarizes the coupled barriers and practical interventions.

\subsection{Fragmented and Heterogeneous Data}

Materials data are heterogeneous by nature. A single material system may require composition, synthesis history, processing parameters, micrographs, spectra, diffraction patterns, electrical properties, thermal properties, mechanical tests, environmental exposure histories, device-level performance, and failure-analysis results. These data differ in scale, format, uncertainty, sampling rate, dimensionality, and vocabulary. A computational database may represent a material through atomic structure and calculated properties, while an experimental laboratory may describe the same material through batch records, furnace conditions, impurity assays, images, spectra, device processing, and performance under specific operating conditions.

The regional challenge is not the absence of useful data. It is the absence of a common data model that connects data across laboratories, instruments, and application domains. A practical regional model should include five linked layers: sample identity, processing history, characterization records, property and performance measurements, and analysis/modeling workflows. Each layer should include provenance, uncertainty, and versioning. For device-oriented materials, the model should also capture architecture, fabrication sequence, operating conditions, and failure modes. FAIR principles provide the foundation~\cite{Wilkinson2016}, while materials-specific infrastructures such as the Materials Project, AFLOW, NOMAD, JARVIS, matminer, and OPTIMADE show how organized data and interoperable interfaces can support reuse and benchmarking~\cite{Jain2013,Curtarolo2013,Draxl2019,Choudhary2020,Ward2018,Andersen2021}.

\subsection{Weak Translation Between Algorithms and Laboratories}

Many materials-informatics studies perform well on benchmark datasets but do not directly improve laboratory decisions. Training data may not cover the relevant process window; measured properties may vary across instruments; small datasets may produce overconfident predictions; and optimization objectives may omit cost, safety, equipment availability, processing time, manufacturability, or qualification requirements. For regional laboratories, the most important questions are often not simply ``which material has the best property?'' but rather ``which process condition should be tested next?'' or ``which failure mechanism limits yield?''

The region should therefore emphasize physics-informed, uncertainty-aware, and decision-centered machine learning. Bayesian optimization and active learning are useful because they use uncertainty to choose informative experiments~\cite{Lookman2019}. Closed-loop and autonomous-laboratory demonstrations show the value of integrating models, measurements, and decision rules~\cite{Tabor2018,Kusne2020,Szymanski2023}. However, regional workflows should initially focus on semi-closed-loop implementation: human-executed experiments guided by model recommendations, standardized metadata, and reproducible analysis notebooks. Success should be measured by improved experimental decisions, such as fewer failed trials, shorter tuning time, higher process yield, reduced characterization cost, or better lifetime prediction.

\subsection{Uneven Access to Infrastructure and Technical Staff}

Large research universities and national laboratories often have high-performance computing, cloud infrastructure, data engineers, instrument scientists, automation specialists, and research software engineers. Smaller institutions may have strong undergraduate programs, local industry relationships, specialized instruments, or field-relevant test environments, but lack the staff and infrastructure needed to convert laboratory data into reusable materials-data products. If data-centric materials science depends only on institutions that already have advanced cyberinfrastructure, it will widen regional disparities.

A regional strategy should therefore provide shared infrastructure and shared expertise. Practical interventions include regional data stewards, research software engineers, instrument-data specialists, shared cloud notebooks, containerized workflows, common metadata templates, secure storage tiers, and roving technical support. Full robotic autonomy is not required at the beginning. Lower-cost steps such as electronic laboratory notebooks, barcode or QR-code sample tracking, automated environmental logging, scripted analysis pipelines, and machine-readable batch records can immediately improve reproducibility and prepare laboratories for higher levels of automation.

\subsection{Workforce Gap at the Materials--Data Interface}

The region needs people who understand both materials science and data science. Students must learn how sample preparation, calibration, instrument drift, microstructure, impurities, processing history, and environmental exposure affect measured properties. They must also learn coding, statistics, uncertainty quantification, data management, visualization, machine learning, and responsible artificial intelligence. National workforce discussions reinforce this need. The MGI strategic plan identifies workforce development as a central component of materials innovation~\cite{MGI2021}, and broader data-science education guidance emphasizes the integration of computational, statistical, mathematical, and domain-specific reasoning~\cite{NASEM2018DataScience}.

A regional workforce program should therefore be modular and hands-on. Shared short courses and stackable credentials could cover materials data management, Python and scientific computing, experimental design, uncertainty quantification, electronic laboratory notebooks, FAIR data practices, physics-informed machine learning, digital twins, cybersecurity, and responsible AI. Students should curate real datasets, reproduce published models, connect instrument outputs to analysis notebooks, and complete team-based projects with industry or national-laboratory mentors.

\subsection{Weak Incentives for Sharing and Reuse}

Researchers are often rewarded for publications and grants, not for metadata quality, reusable code, curated datasets, negative results, or maintained software. Industry partners may also have legitimate concerns about intellectual property, export control, competitive advantage, and data security. A successful ecosystem must make data sharing professionally valuable, technically practical, and legally safe.

Data citation is one mechanism for aligning incentives. The Joint Declaration of Data Citation Principles argues that data should be treated as legitimate, citable research products~\cite{Force11DataCitation2014}, and practical repository guidance emphasizes persistent identifiers, versioning, landing pages, metadata, and citation support~\cite{Fenner2019}. At the same time, not all data can be immediately open. A regional commons should support open public datasets, embargoed academic datasets, consortium-only datasets, controlled-access datasets, and protected industry datasets. The key distinction is that openness and interoperability are not the same. Even protected datasets can use common metadata, sample identifiers, provenance standards, and model documentation so that they remain reusable within approved access tiers.

\begin{table}[htp!]
\centering
\small
\caption{Strategic barriers, practical consequences, and regional interventions for data-centric materials science.}
\label{tab:barriers_interventions}
\begin{tabularx}{\linewidth}{p{0.22\linewidth}Y Y}
\toprule
\textbf{Barrier} & \textbf{Consequence} & \textbf{Regional intervention} \\
\midrule
Fragmented data & Difficult cross-laboratory reuse; limited machine learning; poor provenance & FAIR metadata templates; persistent sample identifiers; controlled vocabularies; interoperable repositories; versioned workflows \\
\addlinespace
Weak algorithm--laboratory translation & Models perform well on benchmarks but fail to improve experimental decisions & Decision-centered metrics; uncertainty-aware models; physics-informed optimization; semi-closed-loop experimental campaigns \\
\addlinespace
Uneven access to infrastructure and staff & Smaller institutions cannot fully participate; duplicated effort; inconsistent data quality & Shared cyberinfrastructure; regional data stewards; cloud notebooks; common templates; cross-institution staff support \\
\addlinespace
Workforce gap & Students can code without understanding materials context, or understand materials without data fluency & Cross-disciplinary modules; hands-on datasets; internships; staff training; integration of data skills into laboratory courses \\
\addlinespace
Weak incentives for sharing and reuse & Valuable data remain local; negative results disappear; software and datasets receive little credit & Data citation; persistent identifiers; tiered access; governance policies; recognition of datasets and software as scholarly outputs \\
\bottomrule
\end{tabularx}
\end{table}

\section{Roadmap for Regional Leadership}
\label{sec:roadmap}

A Great Plains data-centric materials science ecosystem should be built through a staged roadmap that is ambitious enough to create regional identity but practical enough to produce early deliverables. The roadmap proposed here has five actions: establishing a consortium, building a FAIR-aligned data commons, developing shared closed-loop workflows, creating stackable workforce programs, and aligning funding, governance, and translation metrics. These actions are consistent with the national direction set by the MGI, NSF Materials Innovation Platforms, and NSF cyberinfrastructure investments~\cite{MGI2021,NSFMIP2025,NSFCSSI2022}. Table~\ref{tab:roadmap_actions} identifies lead actors, early deliverables, and decision metrics for these actions; Table~\ref{tab:implementation_roadmap} later places the same actions on a staged timeline.

The key design principle is that infrastructure should be created through use. A data commons, workforce program, or consortium will succeed only if it is connected to real materials problems and produces visible value for researchers, students, industry partners, and funding agencies. Each action should therefore be tied to pilot projects that produce reusable artifacts and decision evidence rather than planning documents alone.

\begin{table}[htp!]
\centering
\small
\caption{Operational responsibilities and metrics for the five roadmap actions. This table defines who acts and how progress is measured; Table~\ref{tab:implementation_roadmap} translates the same actions into time-phased milestones.}
\label{tab:roadmap_actions}
\begin{tabularx}{\linewidth}{p{0.20\linewidth}Y Y Y}
\toprule
\textbf{Action} & \textbf{Lead actors} & \textbf{Early deliverables} & \textbf{Decision metrics} \\
\midrule
Consortium and governance & Steering committee, institutional leads, industry and national-lab advisors & Charter; working groups; pilot-selection criteria; data-use principles & Number of participating institutions; active pilots; signed data-sharing or collaboration agreements \\
\addlinespace
FAIR data commons & Data stewards, cyberinfrastructure staff, laboratory leads & Minimum metadata templates; sample IDs; repository prototype; access tiers; dataset citation guidance & Fraction of pilot data with complete provenance; time from experiment to reusable dataset; dataset reuse by non-originating groups \\
\addlinespace
Closed-loop workflows & Domain scientists, modelers, instrument scientists, students & Semi-closed-loop campaigns; active-learning notebooks; uncertainty-aware models; benchmark baselines & Reduction in experimental cycles, tuning time, prediction error, or failed trials relative to expert-only practice \\
\addlinespace
Stackable workforce programs & Faculty, staff, workforce partners, industry mentors & Short courses; certificates; shared modules; internships; staff training & Students and staff trained; modules adopted across institutions; internship placements; employer feedback \\
\addlinespace
Funding and translation metrics & Consortium leadership, state agencies, federal-program leads, industry partners & Proposal pipeline; industry memberships; protected-data workflows; annual evaluation & Sustained external funding; cross-institution publications; prototypes, patents, licenses, startups, or deployed materials solutions \\
\bottomrule
\end{tabularx}
\end{table}

\subsection{Action 1: Establish a Great Plains Materials Data and Discovery Consortium}

The first step is to create a lightweight but durable Great Plains Materials Data and Discovery Consortium that connects universities, national laboratories, industry partners, state agencies, regional cyberinfrastructure organizations, and workforce-development programs. The consortium should not function merely as a mailing list or annual workshop. It should have a written charter, working groups, shared deliverables, and annual review. Initial working groups should include data standards and governance, cyberinfrastructure and software, experimental workflows and instrumentation, workforce and curriculum, and industry translation.

The consortium should begin with a small number of use-inspired pilot themes where early success is plausible. Candidate pilots include high-purity semiconductor and detector materials, quantum and cryogenic materials, corrosion-resistant coatings, polymers and composites for agriculture and manufacturing, battery and energy-storage materials, and biomaterials or sensor platforms. Each pilot should produce scientific results together with a limited set of reusable outputs, such as sample-tracking practices, workflow documentation, analysis notebooks, and training examples.

\subsection{Action 2: Build a FAIR-Aligned Regional Materials Data Commons}

The data commons should be designed as active research infrastructure rather than a passive archive. It should support persistent sample identifiers, dataset versioning, machine-readable metadata, instrument logs, electronic laboratory notebooks, analysis notebooks, uncertainty annotations, access control, and citation of datasets and software. The first phase should define minimum metadata templates for high-value workflows such as crystal growth, thin-film deposition, microscopy, spectroscopy, electrical measurements, mechanical testing, electrochemical cycling, environmental exposure, and cryogenic device testing.

The commons should build on existing materials-data practices where possible. Interoperability efforts such as OPTIMADE demonstrate the value of common interfaces for distributed databases~\cite{Andersen2021}, while tools such as matminer demonstrate the value of reusable feature generation and benchmarking workflows~\cite{Ward2018}. A regional commons should extend these principles toward experimental, processing-rich, device-relevant, and field-relevant datasets.

To make this infrastructure concrete, the commons should be organized as a layered architecture rather than a single file repository. Table~\ref{tab:data_commons_architecture} summarizes the minimum technical components needed for early pilots and later scale-up. The same architecture can support open academic datasets, embargoed project data, controlled-access biomedical or export-sensitive data, and protected industry data while preserving common identifiers, metadata, provenance, and documentation.

\begin{table}[htp!]
\centering
\small
\caption{Proposed technical architecture for a FAIR-aligned regional materials data commons. The emphasis is on practical components that can be implemented through pilot projects and then generalized across materials domains.}
\label{tab:data_commons_architecture}
\begin{tabularx}{\linewidth}{p{0.20\linewidth}Y Y}
\toprule
\textbf{Architecture layer} & \textbf{Core function} & \textbf{Representative implementation element} \\
\midrule
Data layer & Preserve raw, curated, derived, and model-ready data without losing provenance & Raw instrument files; curated tables; feature matrices; model-ready benchmark splits \\
\addlinespace
Identity layer & Link materials across synthesis, processing, characterization, modeling, and publications & Persistent sample IDs; dataset digital object identifiers; ORCID and institutional identifiers where appropriate \\
\addlinespace
Metadata layer & Make experimental context machine-readable and reusable & Processing parameters; instrument settings; calibration records; uncertainty estimates; environmental and operating conditions \\
\addlinespace
Access layer & Balance openness, intellectual property, compliance, and reuse & Public, embargoed, consortium-only, controlled-access, and industry-protected data tiers \\
\addlinespace
Workflow layer & Enable reproducible analysis and model retraining & Versioned notebooks; containers; workflow recipes; model cards; datasheets; software environments \\
\addlinespace
Interoperability layer & Connect local data to national resources and cross-institution tools & Controlled vocabularies; schema mappings; application programming interfaces; OPTIMADE-compatible concepts where applicable \\
\bottomrule
\end{tabularx}
\end{table}

\subsection{Action 3: Develop Shared Closed-Loop Experimental Workflows}

The region should move from isolated datasets toward workflows in which models recommend experiments and experiments improve models. Full autonomy is not required at the beginning. A practical first step is a semi-closed-loop workflow in which active learning or Bayesian optimization recommends the next synthesis, processing, or characterization step; researchers execute the experiment; standardized data files and metadata are captured; analysis notebooks update the dataset; and the model is retrained for the next decision cycle.

Pilot workflows should be chosen based on measurable decision value. Strong pilots should have a defined design space, a measurable target property, a manageable experiment cycle time, and a baseline for comparison. Success should be measured by improvement over expert-only or one-factor-at-a-time experimentation, including fewer experimental cycles, improved yield, faster down-selection, lower characterization cost, reduced prediction error, or improved reproducibility.

\subsection{Action 4: Create Stackable Workforce Programs}

Workforce development should serve undergraduates, graduate students, technicians, postdoctoral researchers, faculty, and industry professionals. A regional curriculum could include Python for materials data, statistics and uncertainty, electronic laboratory notebooks, FAIR data, experimental design, machine learning for structure--property relationships, high-throughput experimentation, autonomous-laboratory concepts, digital twins, data visualization, cybersecurity, and responsible AI. These modules should be embeddable in existing courses, offered as short courses, combined into certificates, or used in summer schools.

The training should be hands-on. Students should curate real datasets, reproduce published models, connect instrument outputs to analysis notebooks, build metadata templates, perform uncertainty analysis, and complete team-based projects with industry or national-laboratory mentors. Technical staff should be included explicitly because research software engineers, data stewards, instrument scientists, laboratory managers, cyberinfrastructure specialists, and automation technicians are essential for sustaining reusable infrastructure.

\subsection{Action 5: Align Funding, Governance, and Translation Metrics}

A sustainable regional effort will require diversified support from federal agencies, state economic-development programs, industry memberships, philanthropic organizations, and institutional commitments. Federal programs related to MGI, Materials Innovation Platforms, cyberinfrastructure, integrated data systems, energy materials, advanced manufacturing, quantum information science, biotechnology, and workforce training provide natural alignment~\cite{MGI2021,NSFMIP2025,NSFCSSI2022}. Proposals should not be built around infrastructure alone; they should show how infrastructure improves materials decisions.

The consortium should track four classes of metrics so that the ecosystem is evaluated by decision value rather than activity alone. \emph{Activity metrics} should include participating institutions, pilot projects, datasets released, users trained, workshops held, metadata templates adopted, and industry partners engaged. \emph{Data-quality metrics} should include the fraction of datasets with complete provenance, calibration records, uncertainty estimates, machine-readable metadata, versioned analysis code, and documented limitations. \emph{Decision-value metrics} should include fewer experimental cycles, shorter tuning time, improved yield, reduced failed trials, lower characterization cost, or lower prediction error on held-out experiments relative to a predeclared baseline. \emph{Translation metrics} should include cross-institution publications, external proposals, industry-funded projects, prototypes, patents, licenses, startups, deployed materials solutions, and workforce placement in materials-data roles.

\section{Implementation Plan and Milestones}
\label{sec:implementation_plan}

Table~\ref{tab:implementation_roadmap} places the roadmap actions on a staged timeline. The four phases are a timing structure, not a separate set of priorities: the near-term phase emphasizes Action 1 and the minimum common practices required for Action 2; the mid-term phase emphasizes Actions 2 and 3; the growth phase scales Actions 2 and 3 while adding Actions 4 and 5; and the long-term phase institutionalizes governance, translation pathways, and sustained funding. The phases are intentionally overlapping: early pilots generate the technical, workforce, and governance lessons needed for later scale-up. Each phase should therefore produce visible evidence of reuse and decision value rather than only planning documents.

The milestone sequence is designed to move from coordination to demonstration and then to scale. In the near term, the consortium selects pilots and establishes the minimum common practices needed for comparable data. In the mid term, those practices are tested in semi-closed-loop campaigns where model recommendations, laboratory execution, and curated results are evaluated against predeclared baselines. The growth and long-term phases then expand the practices that worked across additional domains, institutions, and partners. The HPGe workflow in Section~\ref{sec:ge_pilot_workflow} illustrates how this staged logic can be implemented in one technically demanding but regionally distinctive pilot.

\begin{table}[htp!]
\centering
\footnotesize
\renewcommand{\arraystretch}{1.15}
\caption{Staged implementation roadmap for a Great Plains data-centric materials science ecosystem. Acronyms are defined at first use.}
\label{tab:implementation_roadmap}
\begin{tabularx}{\linewidth}{p{0.16\linewidth}Y Y Y}
\toprule
\textbf{Phase} & \textbf{Primary objectives} & \textbf{Representative deliverables} & \textbf{Success metrics} \\
\midrule

Near term: 0--18 months 
& Organize the network, select pilot themes, define minimum metadata standards, and launch initial training 
& Consortium charter; topical working groups; pilot-project selection; first metadata templates; data-governance principles; shared seminar and training series 
& 5--10 participating institutions; 3--5 pilot datasets; initial industry advisory group; first short-course cohort; draft data-access policy \\

\addlinespace

Mid term: 18--36 months 
& Build the data-commons prototype and demonstrate semi-closed-loop workflows in selected pilot areas 
& Repository prototype; persistent sample identifiers, meaning stable unique IDs that link samples to synthesis, processing, characterization, and modeling records; instrument-to-notebook pipelines; active-learning pilot experiments, in which model uncertainty guides the next measurement or synthesis step; shared analysis notebooks; industry project templates 
& Reduction in experimental cycles relative to baseline; at least 20 reusable datasets; multiple cross-institution publications or proposals; measurable use of common metadata templates \\

\addlinespace

Growth phase: 3--5 years 
& Expand infrastructure, staff support, workforce programs, protected data tiers, and external partnerships 
& Shared data-steward and research-software support; stackable certificates; expanded cyberinfrastructure; protected data tiers for industry, biomedical, or controlled-access datasets; proposal pipeline for NSF Materials Innovation Platforms, NSF cyberinfrastructure and data-infrastructure programs, U.S. Department of Energy (DOE), and state programs 
& Sustained external funding; 100+ trained participants; measurable industry engagement; externally reviewed datasets; reusable software tools; multiple active closed-loop workflows \\

\addlinespace

Long term: 5+ years 
& Establish national visibility, durable governance, and translation pathways for regional materials innovation 
& Regional center or institute; autonomous or semi-autonomous laboratory nodes; national-lab and user-facility partnerships; commercialization and workforce-placement pathways 
& National recognition; deployed materials solutions; patents, licenses, or startups; continuing industry and federal investment; durable cross-state governance \\

\bottomrule
\end{tabularx}
\end{table}

Annual evaluation by internal participants and external advisors should focus on three questions: whether data are becoming more reusable, whether models are improving real experimental decisions, and whether students, staff, and industry partners gain measurable value from participation.

\section{Representative Pilot Workflow: High-Purity Germanium Detector and Quantum Materials}
\label{sec:ge_pilot_workflow}

\begin{figure}[htp!]
    \centering
    \includegraphics[width=0.95\textwidth]{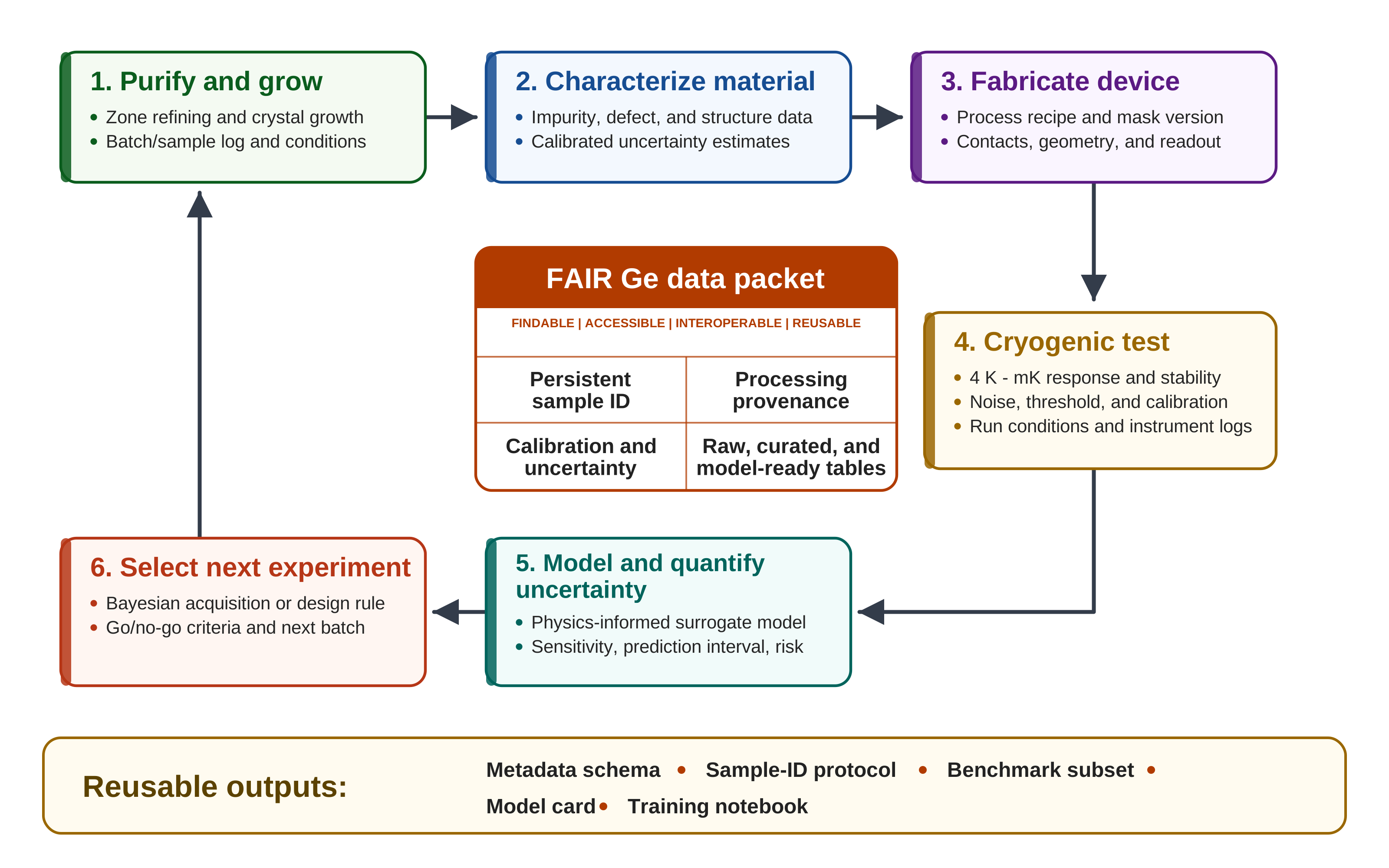}
    \caption{Example high-purity germanium closed-loop pilot workflow for data-centric materials innovation. The workflow links purification and zone refining, Czochralski growth, material characterization, device fabrication, cryogenic testing, uncertainty-aware modeling, and next-experiment selection. It shows how common sample identifiers, processing metadata, calibration records, uncertainty estimates, and analysis notebooks can convert isolated Ge growth and detector measurements into a reusable closed-loop learning system.}
    \label{fig:ge_closed_loop_pilot}
\end{figure}

A concrete pilot is essential because the roadmap will be credible only if it produces early, measurable value. High-purity germanium (HPGe) detector and quantum-materials development is a strong first pilot because it combines regional distinctiveness, experimental complexity, device-level qualification, cryogenic measurement, and national relevance to rare-event detection and semiconductor quantum technologies~\cite{Wang2015,Abgrall2021LEGEND,Watzinger2018,Hendrickx2024}. It also illustrates the manuscript's central argument: the Great Plains can lead by generating processing-rich experimental data that are difficult to obtain from national computational materials databases. Figure~\ref{fig:ge_closed_loop_pilot} illustrates the pilot logic by linking purification, growth, characterization, fabrication, cryogenic testing, modeling, and next-experiment selection.

The pilot objective is to connect the full chain from raw material purification to device performance. The workflow would assign persistent identifiers to each material batch, zone-refined bar, crystal boule, wafer, processed device, and cryogenic test. Each identifier would link to a minimum metadata record covering purification history, growth conditions, impurity assays, carrier concentration, mobility, dislocation density, surface treatment, contact deposition, passivation, packaging, measurement environment, and device-level performance. The same data structure can support detector-grade Ge, Ge-based quantum devices, and related cryogenic semiconductor platforms. Table~\ref{tab:ge_pilot_workflow} summarizes the proposed high-purity Ge pilot workflow, including representative data layers, reusable outputs, and decision-oriented metrics.

\begin{table}[htp!]
\centering
\small
\caption{Concrete high-purity Ge pilot workflow. This pilot converts regional experimental strengths into reusable data products, benchmark models, training modules, and decision metrics.}
\label{tab:ge_pilot_workflow}
\begin{tabularx}{\linewidth}{p{0.18\linewidth}Y Y}
\toprule
\textbf{Workflow layer} & \textbf{Data or activity} & \textbf{Reusable output or metric} \\
\midrule
Materials input & Raw Ge source, purification route, zone-refining pass history, impurity assays, bar position, mass balance & Persistent batch IDs; impurity and purification metadata template; benchmark dataset for impurity reduction and axial variation \\
\addlinespace
Crystal growth & Pulling rate, seed orientation, rotation, thermal gradient, atmosphere, melt history, growth interruptions, boule geometry & Growth-to-crystal metadata template; model features for detector-grade yield and carrier concentration prediction \\
\addlinespace
Characterization & Hall mobility, carrier concentration, resistivity, dislocation density, surface state indicators, microscopy, spectroscopy, uncertainty estimates & Standard analysis notebooks; calibration records; cross-laboratory comparison of measurement uncertainty \\
\addlinespace
Device processing & Wafer location, etch, passivation, lithography, contact deposition, anneal, packaging, thermal history & Fabrication process record; link between contact recipes and leakage, capacitance, noise, and stability \\
\addlinespace
Cryogenic/device testing & Leakage current, capacitance, energy resolution, threshold, noise spectra, charge collection, temperature, bias history, RF or qubit-relevant metrics & Device-performance benchmark; decision metric such as fewer growth/fabrication cycles per viable device or reduced tuning time \\
\addlinespace
Modeling and decision & Physics-informed regression, Bayesian optimization, active learning, uncertainty quantification, failure-mode classification & Model cards; predicted process windows; recommended next experiments; comparison against expert-only baseline \\
\bottomrule
\end{tabularx}
\end{table}

The first modeling task should be deliberately modest: predict whether a given crystal or wafer region is likely to meet predefined detector-grade or device-grade criteria based on purification, crystal-growth, and characterization metadata. A second task could link contact-processing parameters to leakage current, capacitance, noise, or cryogenic stability. A third task could use active learning to recommend the next growth, contact, or surface-treatment condition to test. These tasks are valuable because they address decisions that laboratories already make and because they can be evaluated against baseline practice.

The pilot should use predeclared metrics. Examples include reduction in the number of growth or fabrication cycles required to obtain a viable device, reduction in time to stable cryogenic operation, improvement in prediction error for carrier concentration or leakage current, increase in the fraction of devices meeting predefined benchmarks, and reduction in missing metadata. The pilot should also produce workforce outputs: a sample-ID tutorial, a Ge growth-to-device metadata module, a cryogenic measurement data-curation exercise, and a reproducible notebook that students can run on a shared dataset. In this way, the pilot becomes both a scientific workflow and a training platform.

This example is intentionally specific, but the structure is transferable. Similar pilot templates could be built for corrosion-resistant coatings, battery electrodes, polymer composites, additive-manufacturing feedstocks, antimicrobial surfaces, and sensor packaging. In each case, the goal is the same: convert regional experimental practice into reusable data, uncertainty-aware models, and improved materials decisions.

\section{Regional Use Cases}
\label{sec:regional_use_cases}

\begin{figure}[htp!]
\centering
\resizebox{0.96\linewidth}{!}{%
\begin{tikzpicture}[
    font=\sffamily\small,
    use/.style={
        rectangle,
        rounded corners=8pt,
        draw=black!60,
        line width=0.85pt,
        align=center,
        text width=6.6cm,
        minimum height=1.15cm,
        inner sep=6pt
    },
    center/.style={
        rectangle,
        rounded corners=8pt,
        draw=blue!50!black,
        line width=1.0pt,
        fill=blue!10,
        align=center,
        text width=4.8cm,
        minimum height=1.2cm,
        inner sep=7pt
    },
    outputbox/.style={
        rectangle,
        rounded corners=8pt,
        draw=red!50!black,
        fill=red!8,
        align=center,
        text width=12.0cm,
        minimum height=1.0cm,
        inner sep=7pt
    },
    arr/.style={-{Latex[length=3mm]}, line width=0.9pt, draw=black!70}
]

\node[use, fill=orange!12] (hpge) at (-4.9,2.0)
    {\textbf{High-purity semiconductors,}\\
     \textbf{detectors, and quantum materials}};

\node[use, fill=green!12] (energy) at (4.9,2.0)
    {\textbf{Energy and environmental materials}};

\node[use, fill=cyan!10] (ag) at (-4.9,-0.9)
    {\textbf{Agriculture, manufacturing,}\\
     \textbf{and composites}};

\node[use, fill=purple!10] (bio) at (4.9,-0.9)
    {\textbf{Biomaterials and health technologies}};

\node[center] (platform) at (0,0.55)
    {\textbf{Regional data-centric}\\
     \textbf{materials platform}};

\node[outputbox] (outputs) at (0,-3.0)
    {\textbf{Reusable outputs:}
    FAIR datasets, metadata templates, benchmark models,
    workflows, trained workforce, and translation metrics};

\draw[arr] (hpge) |- (platform);
\draw[arr] (energy) |- (platform);
\draw[arr] (ag) |- (platform);
\draw[arr] (bio) |- (platform);
\draw[arr] (platform) -- (outputs);

\end{tikzpicture}}
\caption{Representative use cases for a Great Plains data-centric materials science ecosystem. A shared regional platform can link these domains to reusable outputs, workforce training, and translation metrics.}
\label{fig:regional_use_cases}
\end{figure}
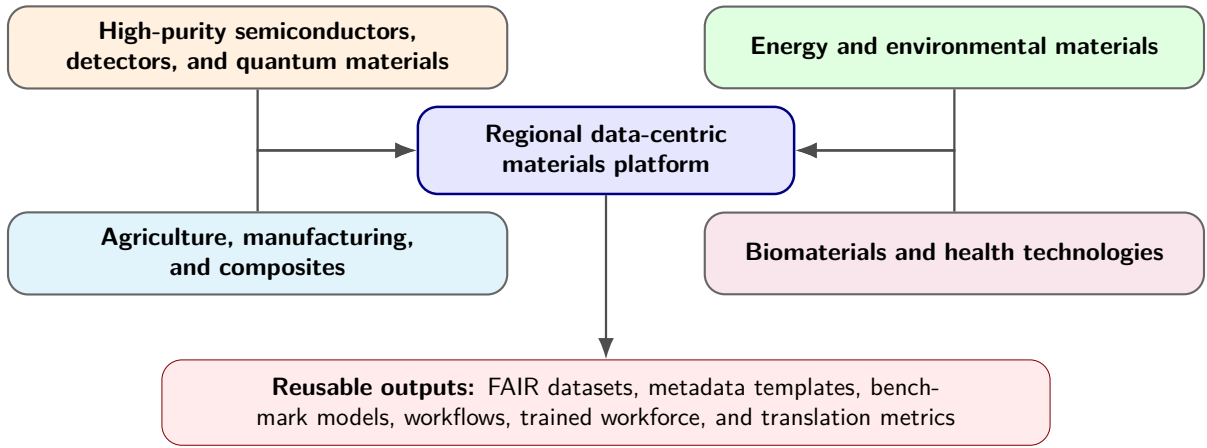

Regional leadership should be organized around use cases where the Great Plains and adjacent interior corridor has distinctive assets, real industrial or societal demand, and measurable opportunities for data-enabled improvement. The strongest use cases are those where performance depends on coupled variables---composition, processing history, microstructure, interfaces, defects, environmental exposure, device architecture, and lifetime---and where better data can improve decisions. Figure~\ref{fig:regional_use_cases} summarizes representative use cases; each should produce reusable datasets, metadata templates, benchmark problems, and training examples that strengthen the broader ecosystem. 

\subsection{High-Purity Semiconductor, Detector, and Quantum Materials}
\label{subsec:semiconductor_detector_quantum}

Because this use case is developed in detail in Section~\ref{sec:ge_pilot_workflow}, we do not repeat the HPGe workflow here. Its broader relevance is methodological: HPGe detectors and Ge-based quantum materials show how a regional ecosystem can connect specialized materials processing, underground or low-background environments, cryogenic testing, reusable metadata, uncertainty-aware modeling, and workforce training in one decision-oriented pilot.

The transferable lesson is that the same data structure---persistent sample identifiers, minimum metadata templates, calibrated measurement records, model documentation, and predeclared decision metrics---can be adapted to other domains where performance depends on processing history, interfaces, defects, and operating conditions. In this sense, the HPGe pilot serves as a proof-of-principle for the broader regional platform rather than a separate or competing use case.

\subsection{Energy and Environmental Materials}
\label{subsec:energy_environmental_materials}

Energy and environmental materials provide broad opportunities for regional collaboration because they connect university research with practical needs in grid reliability, energy storage, catalysis, hydrogen technologies, corrosion, thermal management, water treatment, carbon management, and resource sustainability. Data-centric workflows are especially valuable in these areas because performance often depends on coupled chemistry, microstructure, processing, operating history, and degradation. A battery electrode, for example, cannot be evaluated only by its initial capacity; its value depends on processing route, particle morphology, binder chemistry, electrolyte compatibility, cycling protocol, temperature, rate capability, lifetime, safety, and manufacturability.

Machine learning is already being used to accelerate energy-storage materials discovery, performance prediction, and degradation analysis~\cite{Liu2020BatteryML,Shen2022BatteryML}, but the most useful models require processing and testing metadata, not only final performance values. Regional pilots could connect synthesis conditions, electrode processing, electrochemical cycling, impedance spectra, aging, and failure analysis for batteries, or substrate preparation, coating formulation, exposure history, and failure mode for corrosion-resistant coatings. These datasets would be smaller than large computational databases, but more valuable for deployment because they encode the practical variables that determine lifetime and reliability.

\subsection{Agriculture, Manufacturing, and Composites}
\label{subsec:agriculture_manufacturing_composites}

Agriculture and manufacturing create practical materials problems that are often underrepresented in national materials-informatics datasets, including wear-resistant components, polymer and composite durability, sensor packaging, cold-chain materials, additive-manufacturing feedstocks, corrosion-resistant surfaces, and protective coatings. These problems involve trade-offs among cost, durability, manufacturability, environmental exposure, repairability, supply-chain constraints, and end-user requirements.

Composites, polymers, and additive-manufacturing materials are especially suitable because performance depends on formulation, feedstock, processing history, microstructure, defects, fiber orientation, interfacial bonding, thermal cycling, moisture uptake, and inspection protocol~\cite{MannodiKanakkithodi2021,Chen2021PolymerInformatics,Meng2020AdditiveML}. A regional data-centric approach could link manufacturing centers, mechanical-testing laboratories, microscopy facilities, and industry partners to develop benchmark datasets for feedstock qualification, defect prediction, degradation modeling, and process optimization while training students on field-relevant data.

\subsection{Biomaterials and Health Technologies}
\label{subsec:biomaterials_health}

Regional universities with strengths in biomedical engineering, chemistry, polymer science, imaging, physics, data science, and health systems can contribute to biomaterials and health-technology applications, including antimicrobial surfaces, implant materials, diagnostic sensors, radiation-detection materials, imaging materials, wearable devices, and microfluidic platforms. These topics require students and researchers to integrate materials science with biological constraints, measurement uncertainty, regulatory thinking, privacy, and data ethics.

Biomaterials are a natural fit for data-centric approaches because performance is multidimensional and context-dependent. Cell type, assay conditions, surface preparation, sterilization method, media composition, incubation time, and endpoint definition can all affect measured outcomes~\cite{Terrell2021BiomaterialsML,Fu2025BiomaterialsML}. Antimicrobial surfaces illustrate the metadata challenge: surface chemistry, wettability, roughness, durability, cleaning conditions, and microbial strain all influence measured performance~\cite{Mahanta2021AntimicrobialSurfaces}. For this reason, biomaterials and health-technology pilots should pair standardized metadata with robust governance, including access control, documentation of intended use, validation status, and limitations~\cite{Gebru2021Datasheets,Mitchell2019ModelCards}.

\section{Governance, Data Policy, and Trust}
\label{sec:governance_data_policy_trust}

A regional data-centric materials ecosystem will succeed only if participants trust the rules for data ownership, authorship, intellectual property, embargo periods, access control, cybersecurity, export control, publication timing, and data quality. Trust is especially important in a distributed system that may include universities, national laboratories, small companies, large manufacturers, state agencies, students, and external user facilities. These groups have different incentives and constraints, so the governance framework must promote openness where possible, protection where necessary, and clarity in all cases.

The consortium should adopt a tiered data policy. Public data should be released with persistent identifiers, clear licenses, citation instructions, and sufficient metadata for reuse. Consortium data may be shared among members under agreed terms while datasets are being curated or while multiple groups are contributing to a common benchmark. Industry data may require protected repositories, nondisclosure agreements, restricted access, or aggregation before release. Sensitive data should be handled according to applicable institutional, cybersecurity, export-control, and sponsor policies. The goal is not to force all data into one openness model, but to ensure that all data are managed intentionally. Table~\ref{tab:governance_model} summarizes the proposed governance structure, including data-access policies, quality-control practices, attribution mechanisms, and responsible AI oversight.

\begin{table}[htp!]
\centering
\small
\caption{Operational governance model for a regional materials-data ecosystem.}
\label{tab:governance_model}
\begin{tabularx}{\linewidth}{p{0.20\linewidth}Y Y}
\toprule
\textbf{Governance issue} & \textbf{Proposed policy} & \textbf{Responsible body} \\
\midrule
Dataset access & Use tiered access: public, embargoed, consortium-only, controlled-access, and industry-protected datasets & Data governance working group with institutional compliance offices \\
\addlinespace
Intellectual property and publication timing & Define data-use agreements, embargo periods, authorship expectations, and publication review procedures before data collection begins & Consortium steering committee and participating institutions \\
\addlinespace
Dataset quality & Require minimum metadata, provenance, calibration records, uncertainty estimates, versioning, and known-limitations statements & Data stewards, instrument leads, and pilot-project leads \\
\addlinespace
Model documentation & Use adapted datasheets and model cards to record intended use, training domain, validation data, uncertainty, extrapolation limits, and failure cases & Modeling and software working group \\
\addlinespace
Credit and citation & Assign persistent identifiers to datasets and workflows; require dataset and software citation in publications and reports & Data commons team, journal authors, and consortium evaluation committee \\
\addlinespace
Responsible AI use & Require human review, source traceability, code reproducibility, and disclosure of AI-assisted data curation or analysis when appropriate & Responsible AI and governance subcommittee \\
\bottomrule
\end{tabularx}
\end{table}

Trust also requires quality control. The data commons should distinguish among raw data, curated data, derived features, model-ready tables, and model predictions, while recording provenance, calibration, uncertainty, analysis code, software version, and known limitations~\cite{Wilkinson2016,Ghiringhelli2023Metadata}. Dataset and model documentation should therefore be standard practice. Datasheets and model cards can be adapted to materials science by documenting composition space, processing conditions, measurement methods, intended use, validation domain, uncertainty, extrapolation limits, and known failure cases~\cite{Gebru2021Datasheets,Mitchell2019ModelCards}.

Data citation and responsible AI use should also be built into the governance model. Persistent identifiers, versioned releases, and required dataset or software citation give contributors scholarly credit~\cite{Force11DataCitation2014,Fenner2019}. When generative AI or large language models are used for literature synthesis, code generation, metadata extraction, or experimental planning, the consortium should require human review, source traceability, reproducibility, and disclosure where appropriate. Scientific trust depends on conclusions that can be traced back to validated data and reproducible workflows, not simply on the fluency of AI-generated outputs.

\section{Implementation Risks and Mitigation Strategies}
\label{sec:risks_mitigation}

The roadmap is intentionally practical, but implementation risks should be addressed from the beginning. A regional ecosystem can fail if data quality remains inconsistent, if common metadata become too burdensome, if industry partners do not trust the access rules, if smaller institutions lack technical support, or if machine-learning models do not improve decisions beyond expert-only practice. These risks are manageable if they are treated as design constraints rather than afterthoughts.

\begin{table}[htp!]
\centering
\small
\caption{Implementation risks and mitigation strategies for a distributed regional materials-data ecosystem.}
\label{tab:risks_mitigation}
\begin{tabularx}{\linewidth}{p{0.26\linewidth}Y Y}
\toprule
\textbf{Risk} & \textbf{Potential consequence} & \textbf{Mitigation strategy} \\
\midrule
Inconsistent data quality across laboratories & Models trained on pooled data become unreliable or difficult to interpret & Start with minimum viable metadata, calibration records, uncertainty fields, and annual dataset audits \\
\addlinespace
Metadata burden slows participation & Researchers view FAIR practices as administrative overhead & Use pilot-derived templates, automated instrument-to-notebook pipelines, QR/barcode sample tracking, and data-steward support \\
\addlinespace
Industry reluctance to share process data & High-value deployment data remain outside the commons & Provide tiered access, nondisclosure agreements, aggregation options, embargo periods, and clear intellectual-property rules \\
\addlinespace
Unequal technical capacity among institutions & Smaller institutions contribute less, widening regional disparities & Fund shared data stewards, research software engineers, cloud notebooks, and roving technical support \\
\addlinespace
Weak model-to-laboratory value & AI demonstrations remain disconnected from experimental decisions & Require predeclared baselines, uncertainty estimates, decision metrics, and post-campaign comparison with expert-only practice \\
\addlinespace
Short-lived pilot enthusiasm & Infrastructure is not sustained after initial projects & Tie infrastructure to reusable artifacts, workforce modules, industry value, external proposals, and annual external review \\
\bottomrule
\end{tabularx}
\end{table}

The most important mitigation is to build infrastructure through use. Each pilot should define a manageable data model, a baseline decision process, an access policy, and a small set of success metrics before data collection begins. This approach keeps the commons useful to laboratories while creating evidence for larger center-scale investment.

\section{Discussion}
\label{sec:discussion}

The Great Plains region faces a strategic choice: it can remain a set of strong but loosely connected institutions, or it can organize a network that turns geographic dispersion into a strength. A networked model is well matched to modern materials research because discovery and deployment increasingly depend on distributed synthesis, characterization, computation, user facilities, data engineering, workforce development, and application testing.

The proposed model differs from existing national materials-data resources in its emphasis on experimental, processing-rich, application-relevant data. Early materials informatics was driven by high-throughput computation and open databases such as AFLOW, the Materials Project, NOMAD, and JARVIS~\cite{Curtarolo2013,Jain2013,Draxl2019,Choudhary2020}; recent work emphasizes active learning, autonomous experimentation, and closed-loop workflows~\cite{Tabor2018,Kusne2020,Lookman2019}. The Great Plains can contribute to this next stage by generating datasets that capture processing history, environment, device architecture, calibration, uncertainty, failure modes, and lifetime behavior---information often missing from large computational databases but essential for deployment.

The near-term objective should not be a fully autonomous laboratory or a centralized institute. It should be a trusted operating system for regional collaboration that attaches identity, context, access, analysis, and training practices to real pilot projects. If this operating layer reduces experimental cycles, improves reproducibility, and enables cross-institution reuse, it will justify larger investments in autonomous experimentation, shared user facilities, cyberinfrastructure, and national center-level programs.

Regional leadership should therefore be measured by connectivity, reuse, and decision value. A successful Great Plains network would have cross-institution data reuse, models validated beyond the originating laboratory, students trained on real materials-data workflows, industry problems converted into benchmark challenges, and pilots that demonstrably improve materials decisions relative to baseline practice. Such a model could also guide other distributed research corridors seeking to contribute to the next phase of the Materials Genome Initiative~\cite{MGI2021}.

\section{Conclusions}
\label{sec:conclusions}

Data-centric materials science offers a timely opportunity for the Great Plains and adjacent interior research corridor. The region has relevant assets in universities, national laboratories, cyberinfrastructure networks, specialized test environments, and industries that need deployable materials solutions. Its strengths span high-purity semiconductors and detector materials, quantum and cryogenic devices, energy and environmental materials, polymers and composites, agriculture, manufacturing, biomaterials, and health technologies. Its central challenge is coordination: the region needs better-governed data, interoperable workflows, trained people, shared infrastructure, and translation metrics.

This perspective argues that the region can make a distinctive national contribution by specializing in experimental, processing-rich, application-driven datasets and by connecting those datasets to semi-closed-loop workflows. The proposed roadmap begins with pilot-tested governance, technical, and workforce practices and builds toward protected industry data tiers, autonomous experimentation, and national center-level programs.

If implemented with clear baselines, tiered governance, shared technical support, and annual evaluation, this strategy would position the region as a national contributor to materials innovation while supporting regional economic development and workforce retention. The broader lesson is that leadership in data-centric materials science will come from ecosystems that connect domain expertise, trustworthy data, modern computation, and real-world materials challenges. The Great Plains has the ingredients to build such an ecosystem; its opportunity is to organize those ingredients into a durable, trusted, and application-driven network.
\section*{Author contributions}
D.-M. Mei: conceptualization, methodology, project administration, supervision, funding acquisition, and writing--original draft. K. Acharya, C. M. Adhikari, M. Adhikari, S. Aryal, K. Bhatta, S. Bhattarai, N. Budhathoki, D. Chakraborty, S. Chhetri, S. Choudhury, T. A. Chowdhury, B. Cui, S. Dhital, K.-M. Dong, A. Ghasemi, B. D. S. Gurung, T. Mukherjee,  H. Oli, S. A. Panamaldeniya, L. Pandey, A. Prem, M. M. Rana, L.-W. Wang, Y. Yang, M. Zhou, and Q. Zhou: materials-science, chemistry, physics, device, and experimental-use-case input; investigation of platform requirements; and writing--review and editing. B. V. Benson, R. D. Cruz, A. M. Castillo, E. Z. Gnimpieba, R. Gapuz, H. A. Hashim, R. I. Harry, K.-E. Hasin, M. K. Hassanzadeh, M. K. Jha, D. Kim, K.-C. Kong, B. Lama, A. Mahat, N. Maharjan, A. Majeed, M. M. Masud, K. S. Moore, A. Nawaz, R. Pandey, Z. Peng, R. Rizk, C. S. Tadi, G.-L. Yin, C.-X. Yu, and D. Zeng: AI/ML, biomedical, data-infrastructure, education, governance, computational, and translational perspectives; validation of conceptual scope; and writing--review and editing. All authors reviewed the manuscript and approved the submitted version.

\section*{Conflicts of interest}
The authors declare no conflict of interest related to this conceptual manuscript.

\section*{Data availability}
No new experimental datasets were generated in this manuscript. The paper presents a conceptual platform architecture and cites public data resources and peer-reviewed literature.
\section*{Acknowledgments}
This work was supported in part by \textcolor{blue}{NSF OIA 2427805,} NSF OISE 1743790, NSF PHYS 2117774,  NSF PHYS 2310027, NSF OIA 2437416, DOE DE-SC0024519, DE-SC0004768, and a research center supported by the State of South Dakota.

\bibliographystyle{unsrtnat}
\bibliography{DataCentricMaterials_GreatPlains_CiteStyle_References}

\end{document}